\documentclass{article}

\usepackage{graphics}
\usepackage{psfrag}

\psfrag{Ni/N}[tr][][1.4][270]{$\frac{n_{i}}{N}$}
\psfrag{sqrt(s),MeV}[t][][1.4][0]{$\sqrt{s} \mbox{,~MeV}$}
\psfrag{sigma,nb}[br][][1.4][0]{$\sigma\mbox{,~nb}$}
\psfrag{mgg,MeV}[t][][1][0]{$m_{\gamma\gamma}$,~MeV}
\psfrag{l3g}[t][][1][0]{$\chi^{2}_{3\gamma}$}
\psfrag{l3g-l2g}[t][][1][0]{$\chi^{2}_{3\gamma}-\chi^{2}_{2\gamma}$}
\psfrag{ND}[l][l][1][0]{ND \cite{cite:Dolinsky:nd}}
\psfrag{SND99}[l][l][0.8][0]{SND99 \cite{cite:Achasov:phip0g}}
\psfrag{this study}[l][l][1][0]{this study}
\psfrag{sq(s)=782 MeV}[l][l][1][0]{$\sqrt{s}=782$~MeV}
\psfrag{Er, MeV}[l][l][1][0]{$E_{r}$,~MeV}
\psfrag{e(s(s),Er)/e(s(s),0)}[l][l][1][0]{$\varepsilon_{r}(\sqrt{s},E_{r})/\varepsilon_{r}(\sqrt{s},0)$}
\psfrag{dEr=64 MeV}[l][l][1][0]{$\delta{}E_{r}\sim{}64.4$~MeV}
\psfrag{dEr, MeV}[l][l][1][0]{$\delta{}E_{r}$,~MeV}
\psfrag{sqrt(s), MeV}[l][l][1][0]{$\sqrt{s}$,~MeV}

\begin{document}
\title{Experimental study of the \mbox{$ e^{+}e^{-}\rightarrow \pi ^{0}\gamma  $}
process in the energy region \mbox{$ \sqrt{s}=0.60-0.97 $}~GeV.}

\author{M.N.Achasov,
K.I.Beloborodov, 
A.V.Berdugin, \\
A.G.Bogdanchikov,
A.V.Bozhenok,
A.D.Bukin, 
D.A.Bukin, \\
T.V.Dimova,
V.P.Druzhinin,
V.B.Golubev, 
V.N.Ivanchenko, \\
A.A.Korol,
S.V.Koshuba,
E.V.Pakhtusova, 
A.A.Polunin, \\
E.E.Pyata,
A.A.Salnikov,
S.I.Serednyakov, 
V.V.Shary, \\ 
Yu.M.Shatunov,
V.A.Sidorov,
Z.K.Silagadze, 
A.N.Skrinsky, \\
A.G.Skripkin,
Yu.V.Usov,
A.A.Valishev, 
A.V.Vasiljev}



\maketitle{}

\begin{abstract}
Results of the study of the $ e^{+}e^{-}\rightarrow \pi ^{0}\gamma  $
process with SND detector at VEPP-2M collider in the c.m.s. energy
range $ \sqrt{s}=0.60-0.97 $~GeV are presented. Using 36513 selected
events corresponding to a total integrated luminosity of $ 3.4\, pb^{-1} $
the $ e^{+}e^{-}\rightarrow \pi ^{0}\gamma  $ cross section was
measured. The energy dependence of the cross section was analyzed
in the framework of the vector meson dominance model. The data are
well described by a sum of $ \phi ,\omega ,\rho \rightarrow \pi ^{0}\gamma  $
decay contributions with measured decay probabilities: $ Br(\omega \rightarrow \pi ^{0}\gamma )=(9.34\pm 0.15\pm 0.31)\, \% $
and $ Br(\rho ^{0}\rightarrow \pi ^{0}\gamma )=(5.15\pm 1.16\pm 0.73)\times 10^{-4} $.
The $ \rho -\omega  $ relative interference phase is $ \varphi _{\rho \omega }=-10.2\pm 6.5\pm 2.5^{\circ } $. 
\end{abstract}

\section{Introduction}

Cross section of the $ e^{+}e^{-}\rightarrow \pi ^{0}\gamma  $
process at the c.m.s. energies $ \sqrt{s}=0.60-0.97 $~GeV within
the framework of the vector meson dominance model is determined by
radiative decays of light vector mesons $ \rho ^{0}(770) $, $ \omega (782) $,
$ \phi (1020) $. These decays belong to the class of magnetic dipole
transitions and represent major interest for study of quark structure
of vector mesons and for tests of low-energy models of strong interactions,
such as non-relativistic quark model, effective potential models,
etc. \cite{cite:Geffen:thnqm,cite:ODonnell:thvdm,cite:Singer:thcbm,cite:Barik:theff,cite:Zhu:thsum}.
Study of this process allows to improve accuracy of the parameters
of the $ \rho ^{0},\omega \rightarrow \pi ^{0}\gamma  $ decays.

The only previous measurement of the decay $ \rho ^{0}\rightarrow \pi ^{0}\gamma  $
was carried out by ND detector \cite{cite:Dolinsky:nd}: $ Br(\rho ^{0}\rightarrow \pi ^{0}\gamma )=(7.9\pm 2.0)\times 10^{-4} $.
This result agrees with the PDG value for isotopically complementary
channel: $ Br(\rho ^{\pm }\rightarrow \pi ^{\pm }\gamma )=(4.5\pm 0.5)\times 10^{-4} $
\cite{cite:Hagiwara:pdg}. The $ \omega \rightarrow \pi ^{0}\gamma  $
decay was studied in several experiments \cite{cite:Dolinsky:nd,cite:Jacquet:ome,cite:Baldin:1971bv,cite:Benaksas:1972pp,cite:Keyne:1976tj,cite:Aulchenko:2000a}.
The world average decay probability $ Br(\omega \rightarrow \pi ^{0}\gamma )=(8.7\pm 0.4)\% $
\cite{cite:Hagiwara:pdg}. 

In this work we present the study of the $ e^{+}e^{-}\rightarrow \pi ^{0}\gamma  $
process with SND detector at VEPP-2M collider.

\section{Detector and experiment}

The SND detector \cite{cite:Achasov:snd} consists of an electromagnetic
calorimeter, tracking and muon systems. The main part of the detector
is a three-layer spherical electromagnetic calorimeter consisting
of 1600 \mbox{NaI(Tl)} crystals. Total thickness of the calorimeter
for the particles flying from the interaction point is $ 13.4\, X_{0} $.
Total solid angle is $ 90\%\cdot 4\pi  $. The energy resolution
of the calorimeter for photons is $ \sigma _{E}/E\approx 4.2\%/E(\mathrm{GeV})^{1/4} $,
the angular resolution is $ \sigma _{\varphi ,\theta }\approx 0.82^{\circ }/\sqrt{E(\mathrm{GeV})}\oplus 0.63^{\circ } $.

The experiment was carried out at VEPP-2M collider \cite{cite:Koop:vepp}
with SND detector \cite{cite:Achasov:snd}. The data were collected
in March -- July, 1998 \cite{cite:Achasov:exp96-98} at 30 points
in the energy range $ \sqrt{s}=0.60-0.97 $~GeV. The total integrated
luminosity used for the analysis was $ \sim 3.4 $~pb$ ^{-1} $.
The beam energy determination was based on measured magnetic field
in the bending magnets and beam revolution frequency in the collider.
The error of the center of mass energy determination consists of two
parts: 0.1~MeV --- relative accuracy of energy setting for each energy
point and 0.2~MeV --- general energy scale bias common for all points
within the experiment.

\section{Data analysis}

In this work the process $ e^{+}e^{-}\rightarrow \pi ^{0}\gamma  $
was studied in the three-photon final state. The main sources of background
are QED processes $ e^{+}e^{-}\rightarrow 3\gamma  $ and $ e^{+}e^{-}\rightarrow 2\gamma  $
with extra photons of the machine background. Other possible sources
are process $ e^{+}e^{-}\rightarrow \eta \gamma  $, and cosmic
background.

\subsection{Events selection}

For an event to be recorded the SND first level trigger (FLT) required
at least two clusters of hit crystals in the calorimeter and no signals
in neither tracking nor muon systems. The FLT threshold on calorimeter
energy deposition changed with the beam energy, but was always below
$ 0.4\sqrt{s} $.

The reconstructed events were first put through primary selection
which required at least $ 3 $ neutral and no charged particles,
total energy deposition $ E_{tot}>0.65\sqrt{s} $, total momentum
measured by the calorimeter $ P_{tot}<0.3\sqrt{s} $, polar angles
of the two highest energy photons $ 36^{\circ }<\theta _{1,2}<144^{\circ } $,
the polar angle of the third photon (descending order in energy) $ 27^{\circ }<\theta _{3}<153^{\circ } $,
and the energy deposition of this photon $ E_{p3}>0.1\sqrt{s} $.
These conditions select three-photon events, suppress machine background
and two photon annihilation events with additional background clusters
in the calorimeter. As a result $ 52415 $ events were selected
for further analysis.

\begin{figure}[h!]
{\centering \resizebox*{0.45\textwidth}{!}{\includegraphics{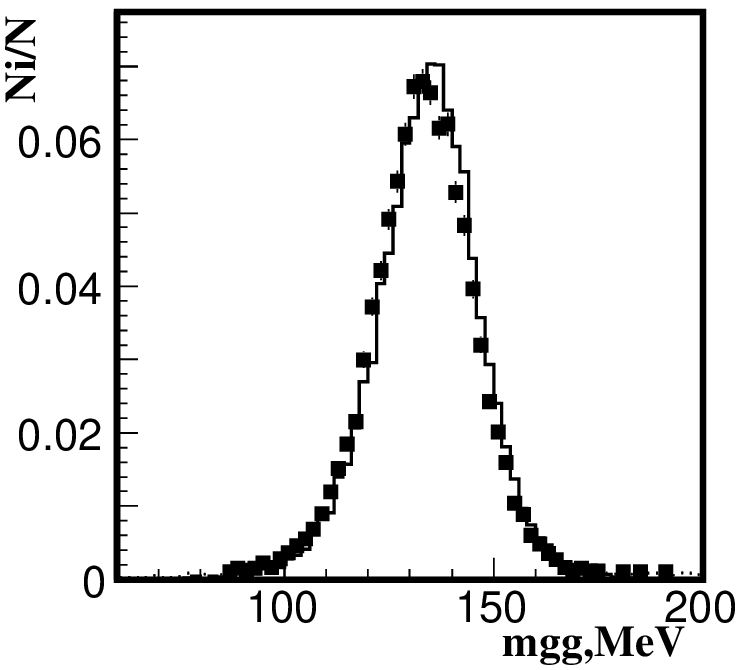}} ~\resizebox*{0.45\textwidth}{!}{\includegraphics{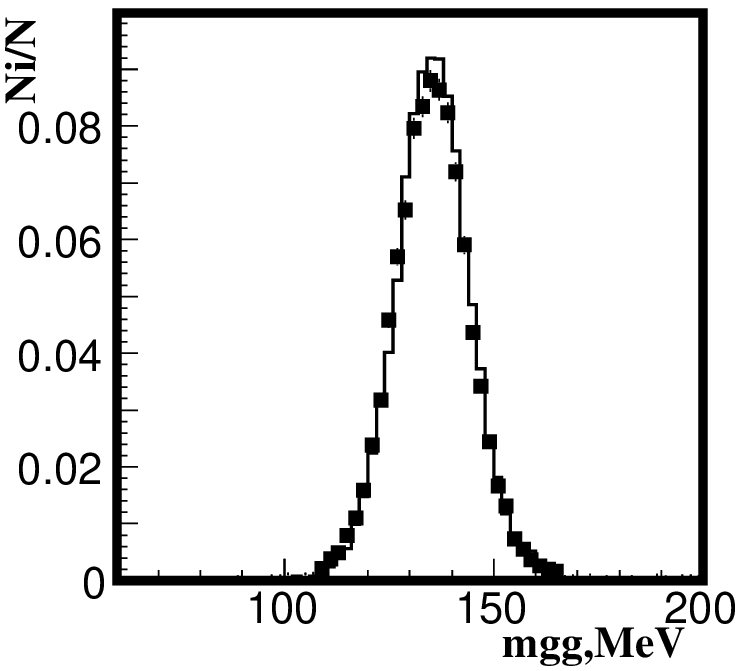}} \par}

\caption{\label{fig:m2g}Invariant mass distribution for photons pairs in
\protect$ e^{+}e^{-}\rightarrow \pi ^{0}\gamma \protect $ events
before (left) and after (right) kinematic fitting. Solid line~---
MC simulation, points~--- data (\protect$ \sqrt{s}=782\protect $~MeV). }
\end{figure}

In order to improve energy and angular resolution for photons the
selected events were kinematically fitted with total energy and momentum
conservation constraints. The fit results are the value of $ \chi _{3\gamma }^{2} $
of the hypothesis and fitted kinematical parameters of the photons.
The kinematic fit improves resolution in invariant mass of photon
pairs from $ \pi ^{0} $ decays from $ 11.2 $~MeV to $ 8.6 $~MeV
(Fig.\ref{fig:m2g}). 

\begin{figure}[h!]
{\centering \resizebox*{0.45\textwidth}{!}{\includegraphics{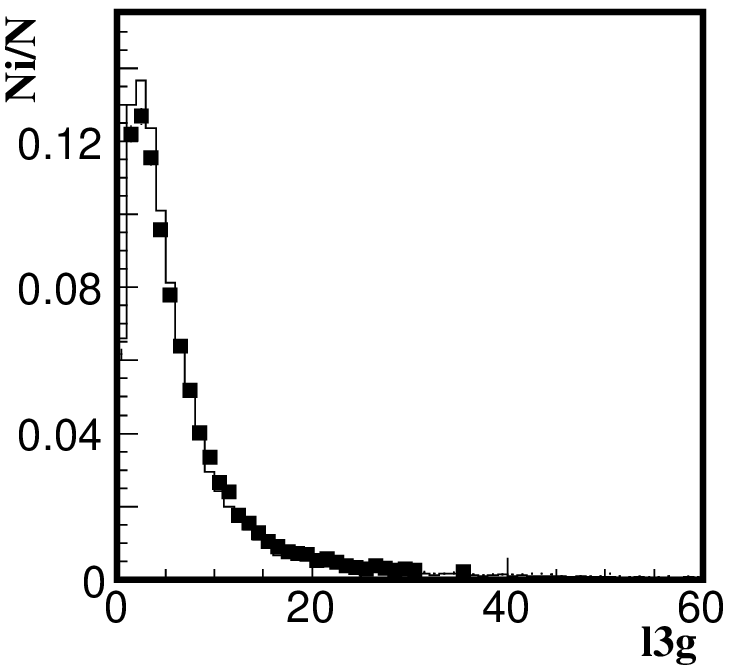}} ~\resizebox*{0.45\textwidth}{!}{\includegraphics{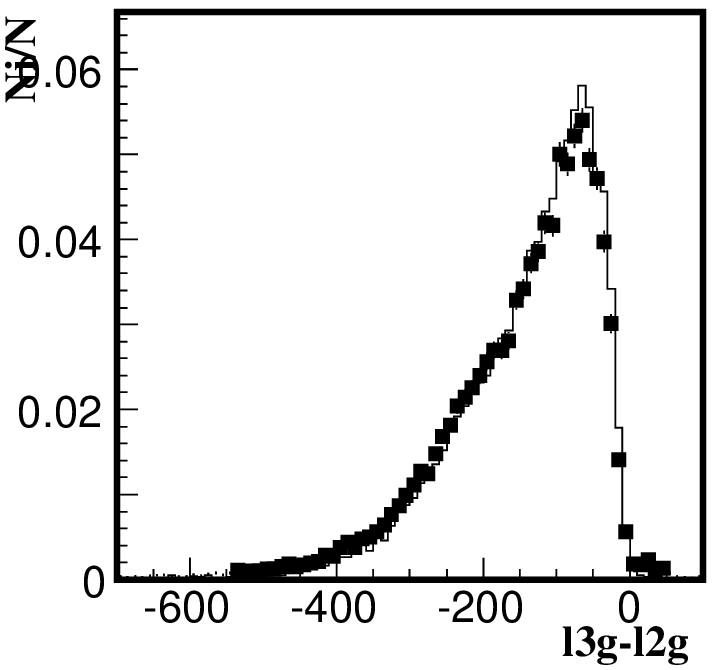}} \par}

\caption{\label{fig:xi2}Distributions over \protect$ \chi ^{2}_{3\gamma }\protect $
(left) and \protect$ \chi ^{2}_{3\gamma }-\chi ^{2}_{2\gamma }\protect $
(right) for class \textbf{\textsl{A}} selection. Solid line~--- MC
simulation, points~--- experiment (\protect$ \sqrt{s}=782\protect $~MeV). }
\end{figure}

The $ \chi ^{2}_{3\gamma } $ distribution is shown in Fig.\ref{fig:xi2}.
For additional suppression of cosmic and machine backgrounds we required
$ \chi _{3\gamma }^{2}<20 $. This cut also implicitly limits maximum
energy of initial state radiation (ISR) photons in the process under
study (Fig.\ref{fig:rceff}). In order to suppress 2-photon annihilation
background, each selected event was kinematically fitted to $ e^{+}e^{-}\rightarrow 2\gamma  $
hypothesis and restriction on calculated $ \chi ^{2}_{2\gamma } $
was applied: $ \chi ^{2}_{3\gamma }-\chi ^{2}_{2\gamma }<0 $. The
$ \chi ^{2}_{3\gamma }-\chi ^{2}_{2\gamma } $ distribution is shown
in Fig.\ref{fig:xi2}. 

The only significant backgrounds to the process under study remaining
after described above cuts are $ e^{+}e^{-}\rightarrow \eta \gamma  $
and QED $ 3\gamma  $ annihilation. In the latter process all kinematically
allowed combinations of the photon energies and angles are present,
so this background cannot be completely eliminated and must be subtracted.
To this end all events, which passed primary selection and kinematic
fit cuts, were divided into two classes: events with $ 108\, \mathrm{MeV}\leq m_{\gamma \gamma }\leq 162\, \mathrm{MeV} $
were assigned to class \textbf{\textsl{A}}, the rest --- to class
\textbf{\textsl{B}}. Here $ m_{\gamma \gamma } $ is an invariant
mass of a photon pair after kinematic fitting (Fig.\ref{fig:m2g}).
The total number of the selected class \textbf{\textsl{A}} events
is 36513. The fraction of $ e^{+}e^{-}\rightarrow \eta \gamma  $
background in this class is less than $ 0.1\% $. For the class
\textbf{\textsl{B}} this fraction is up to $ 5\% $. For calculation
of integrated luminosity the special sample of $ e^{+}e^{-}\rightarrow 2\gamma  $
events (class \textbf{\textsl{C}}) was selected using following criteria:
no charged particles, at least two neutral particles, energy depositions
for two most energetic photons $ E_{p1,2}>0.3\sqrt{s} $, their
polar angles $ 36^{\circ }<\theta _{1,2}<144^{\circ } $, azimuth
acollinearity $ \Delta \varphi _{12}<10^{\circ } $, polar acollinearity
$ \Delta \theta _{12}<25^{\circ } $, event does not belong to classes
\textbf{\textsl{A}} or \textbf{\textsl{B}}. It is necessary to note
significant contribution of $ e^{+}e^{-}\rightarrow \pi ^{0}\gamma  $
events to the class \textbf{\textsl{C}} (up to $ 10\% $ at $ \omega (782) $
resonance).

\subsection{Cross section parameterization }

The $ e^{+}e^{-}\rightarrow \pi ^{0}\gamma  $ cross section in
the framework of VDM can be parameterized as follows \cite{cite:Achasov:freshlook,cite:OConnell:mixing}:
\begin{eqnarray}
\sigma _{\pi ^{0}\gamma }(s) & = & \frac{(4\pi )^{2}\alpha \cdot q(s)^{3}}{3s^{3/2}}\left| \sum _{V=\rho ,\omega ,\phi }\frac{g_{\gamma V}\cdot g_{V\pi ^{0}\gamma }}{D_{V}(s)}+A_{nonres}\right| ^{2},\label{mth:crosspg} \\
D_{V}(s) & = & m_{V}^{2}-s-i\sqrt{s}\Gamma _{V}(s),\\
q(s) & = & \frac{\sqrt{s}}{2}\left( 1-\frac{m^{2}_{\pi ^{0}}}{s}\right) .
\end{eqnarray}
 Here $ g_{\gamma V} $ and $ g_{V\pi ^{0}\gamma } $ are coupling
constants, $ m_{V} $ is the $ V $ resonance mass, $ \Gamma _{V}(s) $
is the energy-dependent width of the resonance taking into account
processes with branching ratios larger than $ 1\% $, $ A_{nonres} $
represents possible non-resonant contribution. Using following formulas
for coupling constants: \begin{eqnarray}
\left| g_{\gamma V}\right|  & = & \sqrt{\frac{m_{V}^{5}}{(4\pi )^{2}\alpha }\Gamma _{V}\sigma _{V}},\label{mth:ggv} \\
\left| g_{V\pi ^{0}\gamma }\right|  & = & \sqrt{\frac{3\Gamma _{V}}{q(m^{2}_{V})^{3}}\frac{\sigma _{V\pi ^{0}\gamma }}{\sigma _{V}}},
\end{eqnarray}
 where $ \sigma _{V} $ and $ \sigma _{V\pi ^{0}\gamma } $ are
cross sections of $ e^{+}e^{-}\rightarrow V $ and $ e^{+}e^{-}\rightarrow V\rightarrow \pi ^{0}\gamma  $
for $ \sqrt{s}=m_{V} $, Eq.(\ref{mth:crosspg}) can be transformed
to the form more suitable for data approximation:\begin{eqnarray}
\sigma _{\pi ^{0}\gamma }(s) & = & \frac{q(s)^{3}}{s^{3/2}}\left| A_{\rho ^{0}\pi ^{0}\gamma }(s)+A_{\omega \pi ^{0}\gamma }(s)+A_{\varphi \pi \gamma }(s)+a_{\pi ^{0}\gamma }\right| ^{2},\\
A_{V}(s) & = & \frac{m_{V}\Gamma _{V}f_{V}(s)}{D_{V}(s)}\sqrt{\frac{m_{V}^{3}}{q(m_{V}^{2})^{3}}\sigma _{V\pi ^{0}\gamma }},
\end{eqnarray}
 where $ a_{\pi ^{0}\gamma } $ is a non-resonant contribution.
We used two different models for description of interference phase
between $ \rho ,\omega \rightarrow \pi ^{0}\gamma  $ decay amplitudes.
For model with energy-independent interference phases, $ f_{\rho ,\varphi }=e^{i\varphi _{\rho ,\varphi }},f_{\omega }\equiv 1 $.
In this case the $ \varphi _{\rho } $ is expected to be $ 0^{\circ } $
for pure $ \rho  $ and $ \omega  $ isotopic states. Electromagnetic
$ \rho  $-$ \omega  $ mixing leads to nonzero value of $ \varphi _{\rho } $.
It can be estimated from $ B(\omega \rightarrow 2\pi ) $: $ \varphi _{\rho }\approx -13^{\circ } $.
The second model is based on mixed propagator approach \cite{cite:Achasov:mixing,cite:OConnell:mixing}:

\begin{eqnarray}
f_{\rho ,\omega }(s)=\frac{r_{\rho ,\omega }(s)}{\left| r_{\rho ,\omega }\left( m_{V}^{2}\right) \right| } & , & f_{\phi }(s)=e^{i\varphi _{\phi }},\label{mth:mixing} \\
r_{\omega }(s) & = & 1+\varepsilon (s)\cdot \left( \frac{\left| g_{\gamma \rho ^{0}}\right| }{\left| g_{\gamma \omega }\right| }+\frac{\left| g_{\rho ^{0}\pi ^{0}\gamma }\right| }{\left| g_{\omega \pi ^{0}\gamma }\right| }\right) ,\\
r_{\rho }(s) & = & 1-\varepsilon (s)\cdot \left( \frac{\left| g_{\gamma \omega }\right| }{\left| g_{\gamma \rho ^{0}}\right| }+\frac{\left| g_{\omega \pi ^{0}\gamma }\right| }{\left| g_{\rho ^{0}\pi ^{0}\gamma }\right| }\right) ,\\
\varepsilon (s) & = & \frac{\Pi _{\rho \omega }}{D_{\omega }(s)-D_{\rho }(s)},\label{mth:mixingl} 
\end{eqnarray}
 where $ \Pi _{\rho \omega } $ is $ \rho -\omega  $ mixing self-energy.

\begin{figure}[h!]
\resizebox*{0.4\textwidth}{!}{\includegraphics{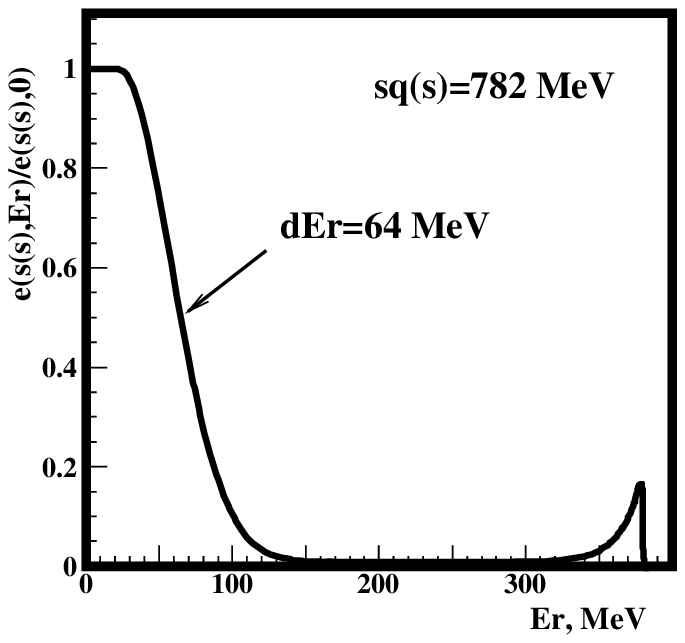}} \resizebox*{0.4\textwidth}{!}{\includegraphics{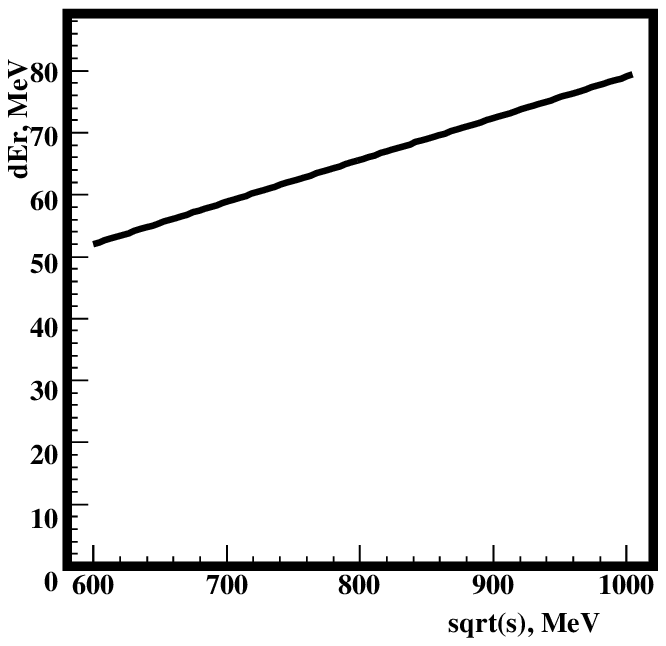}}

\caption{\label{fig:rceff}Detection efficiency as function of the extra photons
energy \protect\protect$ \varepsilon _{r}(\sqrt{s},E_{r})\protect $
for \protect\protect$ \sqrt{s}=782\protect $~MeV~(left) and effective
extra photons energy threshold \protect\protect$ \delta E_{r}\protect $
(right) as function of \protect\protect$ \sqrt{s}\protect $.}
\end{figure}

Detection efficiency for $ e^{+}e^{-}\rightarrow \pi ^{0}\gamma  $
process depends not only on c.m.s. energy $ \sqrt{s} $ but also
on energy of extra photons emitted by initial particles $ E_{r} $.
The detection efficiency $ \varepsilon _{r}(\sqrt{s},E_{r}) $ was
determined by Monte Carlo simulation with ISR taken into account.
The energy and angular distributions of ISR photons were generated
according to Refs. \cite{cite:Kuraev:radcor,cite:Bonneau:radcor}.
The dependence of detection efficiency on $ E_{r} $ approximated
by a smooth function is shown in Fig.\ref{fig:rceff} for $ \sqrt{s}=782 $~MeV.
The noticeable peak in detection efficiency near ISR kinematic limit
corresponds to the case when $ \pi ^{0}\gamma  $ invariant mass
is close to $ m_{\pi ^{0}} $ and ISR photon is emitted at the large
angle and detected. The effective threshold on the ISR photon energy
$ \delta E_{r}(s) $ is determined by $ \chi ^{2} $~restriction
and can be defined as a width at half maximum of the $ \varepsilon _{r}(\sqrt{s},E_{r}) $.
At $ \sqrt{s}=782 $~MeV $ \delta E_{r}\approx 64.4 $. The $ \delta E_{r} $
dependence on $ \sqrt{s} $ is shown in Fig.\ref{fig:rceff}.

The visible cross section of the $ e^{+}e^{-}\rightarrow \pi ^{0}\gamma  $
process was calculated as \cite{cite:Bozhenok:fit}: \begin{eqnarray}
\sigma _{vis}(s) & = & \int _{0}^{\frac{2E_{r,max}}{\sqrt{s}}}\varepsilon _{r}(\sqrt{s},\frac{x\sqrt{s}}{2})F(x,s)\sigma ((1-x)s)dx\label{mth:viscalc} 
\end{eqnarray}
 where $ \sigma (s) $~is cross section, the function $ F(x,s) $
is electron {}``radiator'' function \cite{cite:Kuraev:radcor}.
For data presentation we used traditional form:\begin{eqnarray}
\sigma _{vis}(s) & = & \varepsilon (\sqrt{s})\cdot \beta (\sqrt{s})\cdot \sigma (s)\nonumber \label{mth:visdef} 
\end{eqnarray}
 where $ \varepsilon (\sqrt{s}) $ and $ \beta (\sqrt{s}) $ are
defined as:\begin{eqnarray}
\varepsilon (\sqrt{s}) & \equiv  & \varepsilon _{r}(\sqrt{s},0)\label{mth:viseff} \\
\beta (\sqrt{s}) & \equiv  & \frac{\int _{0}^{\frac{2E_{r,max}}{\sqrt{s}}}\varepsilon _{r}(\sqrt{s},\frac{x\sqrt{s}}{2})F(x,s)\sigma ((1-x)s)dx}{\varepsilon _{r}(\sqrt{s},0)\cdot \sigma (s)}\label{mth:visbeta} 
\end{eqnarray}

For simulation of the background process $ e^{+}e^{-}\rightarrow 3\gamma  $~(QED)
the lowest-order formulas from \cite{cite:Arbuzov:thqed} were used.
Visible cross section calculated using Monte Carlo simulation was
corrected for higher order loop diagrams and soft photons emission
\cite{cite:Kuraev:th3grc} using $ \delta E_{r} $ as an upper limit
of soft photons energy. The correction varied in the range of $ 0.915-0.925 $.
We expect that the accuracy of the calculated $ e^{+}e^{-}\rightarrow 3\gamma  $
visible cross section is not worse than $ 2\% $.

For simulation of the process $ e^{+}e^{-}\rightarrow 2\gamma  $~(QED)
used for luminosity determination the formula from \cite{cite:th2g}
taking into account additional photon emission was used. The accuracy
of the visible cross section determination is estimated as $ 1\% $.

\subsection{Data approximation}

The FIT package \cite{cite:Bozhenok:fit} was used for data fitting.
The fitting was done by means of maximum likelihood method on all
three data sets (classes \textbf{\textsl{A}},\textbf{\textsl{B}},and
\textbf{\textsl{C}}) simultaneously. Expected number of events in
the $ i $-th energy point was calculated as:\begin{eqnarray}
N^{(j)}_{i} & = & IL_{i}\cdot (\sigma _{\pi ^{0}\gamma ,vis}^{(j)}(E_{i})+\sigma _{3\gamma ,vis}^{(j)}(E_{i})+\sigma _{\eta \gamma ,vis}^{(j)}(E_{i}));\, j=A,B;\nonumber \\
IL_{i} & = & \frac{N^{(C)}_{i}}{\sigma _{2\gamma ,vis}^{(C)}(E_{i})+\sigma _{\pi ^{0}\gamma }^{(C)}(E_{i})}\, .\nonumber 
\end{eqnarray}
 Visible hadronic cross sections were calculated according to Eq.(\ref{mth:viscalc})
and corrected for beam energy spread. Because the $ e^{+}e^{-}\rightarrow \pi ^{0}\gamma  $
process gives noticeable contribution to events of luminosity process
$ e^{+}e^{-}\rightarrow 2\gamma  $, the integrated luminosity ($ IL_{i} $)
was recalculated on every iteration step of the minimization. The
accuracy of determination of the c.m.s. energy is worse than the accuracy
of the $ \omega  $-meson mass value. Therefore we introduced the
possible energy scale bias $ \Delta E $ as a free parameter. Other
fit parameters were $ \sigma _{\omega \pi ^{0}\gamma } $, $ \sigma _{\rho \pi ^{0}\gamma } $,
$ a_{\pi ^{0}\gamma } $ , $ \varphi _{\rho } $ or $ \Pi _{\rho \omega } $
depending on approach to phase factor (Eqs.(\ref{mth:crosspg})--(\ref{mth:mixing}))
calculation, and $ k_{3\gamma } $. The $ k_{3\gamma } $ parameter
is a ratio of measured and calculated $ \sigma _{3\gamma } $ cross
sections. Parameters of the process $ e^{+}e^{-}\rightarrow \phi \rightarrow \pi ^{0}\gamma  $
were taken from \cite{cite:Achasov:phip0g}: \begin{eqnarray}
\sigma _{\phi \pi ^{0}\gamma } & = & 5.12\pm 0.39\, \mbox {nb}\label{mth:salnikov} \\
\varphi _{\phi } & = & 158\pm 11^{\circ }
\end{eqnarray}
 For other cross section parameters the world average values \cite{cite:Hagiwara:pdg}
were used.

Data were approximated in four models:

\begin{enumerate}
\item $ \sigma _{\rho ^{0}\pi ^{0}\gamma }=0 $, $ a_{\pi ^{0}\gamma }=0 $, 
\item $ \sigma _{\rho ^{0}\pi ^{0}\gamma } $ and $ \varphi _{\rho ^{0}} $
are free parameters, $ a_{\pi ^{0}\gamma }=0 $, 
\item $ \sigma _{\rho ^{0}\pi ^{0}\gamma } $ is a free parameter, $ a_{\pi ^{0}\gamma }=0 $,
$ \Pi _{\rho \omega } $ is a free parameter, 
\item $ \sigma _{\rho ^{0}\pi ^{0}\gamma } $ is a free parameter, $ a_{\pi ^{0}\gamma } $
is a free real parameter, $ \Pi _{\rho \omega } $ calculated from
$ B(\omega \rightarrow 2\pi ) $. 
\end{enumerate}
Obtained energy scale bias ($ \Delta E $) is $ -0.34\pm 0.08 $~MeV
for all models. It is consistent with our expectations. Found value
of $ k_{3\gamma } $ is $ 98.7\pm 1.3\% $ with $ \frac{\chi ^{2}}{N}=\frac{26}{29} $
shows good agreement between calculated and measured cross sections
of QED $ 3\gamma  $ annihilation. For background subtraction we
used the measured cross section. Other obtained parameters are listed
in the Table \ref{tab:variants}. 
\begin{table}[h!]

\caption{\label{tab:variants}The fitted cross section parameters for different
models. Only statistical errors are shown. }

\centering

\begin{tabular}{|c|c|c|c|c|c|c|}
\hline 
&
 $ \sigma _{\omega \pi ^{0}\gamma } $, nb&
 $ \sigma _{\rho ^{0}\pi ^{0}\gamma } $, nb&
 $ \scriptstyle \varphi _{\rho \omega },^{\circ } $&
 $ \Pi _{\rho \omega } $,MeV$ ^{2} $&
 $ Re\, a_{\pi ^{0}\gamma } $,nb$ ^{\frac{1}{2}} $&
 $ \chi ^{2}/N $\\
\hline
1&
 $ \scriptstyle 176.6\pm 1.4 $&
 $ \scriptstyle 0 $&
&
&
 $ \scriptstyle 0 $&
 $ \scriptstyle 81/28 $\\
\hline
2&
 $ \scriptstyle 155.8\pm 2.7 $&
 $ \scriptstyle 0.58\pm 0.13 $&
 $ \scriptstyle -10.2\pm 6.5 $&
&
 $ \scriptstyle 0 $&
 $ \scriptstyle 21.6/26 $\\
\hline
4&
 $ \scriptstyle 155.9\pm 2.7 $&
 $ \scriptstyle 0.56\pm 0.13 $&
 $ \scriptstyle -9.9\pm 6.5 $$ \scriptstyle ^{1)} $&
$ \scriptstyle -2819\pm 1841 $&
 $ \scriptstyle 0 $&
 $ \scriptstyle 21.9/26 $\\
\hline
3&
 $ \scriptstyle 156.8\pm 2.8 $&
 $ \scriptstyle 0.51\pm 0.13 $&
 $ \scriptstyle -12.8\pm 1.1 $$ \scriptstyle ^{1)} $&
$ \scriptstyle -3676\pm 303 $$ \scriptstyle ^{2)} $&
 $ \scriptstyle -0.13\pm 0.13 $&
 $ \scriptstyle 20.7/25 $ \\
\hline
\end{tabular}~\raggedright \\
 $ ^{1)} $calculated using Eqs.(\ref{mth:mixing})--(\ref{mth:mixingl}) \\
 $ ^{2)} $derived from $ B(\omega \rightarrow 2\pi ) $

\end{table}

Large $ \chi ^{2} $ value for the first model shows that $ e^{+}e^{-}\rightarrow \pi ^{0}\gamma  $
cross section cannot be described only by $ \omega  $ and $ \phi  $
decays contribution. The second model corresponds to an energy independent
$ \rho -\omega  $ interference phase. Obtained value of this phase
$ (-10.2\pm 6.5)^{\circ } $ is in agreement with expectation from
electromagnetic $ \rho -\omega  $ mixing $ \varphi _{\rho }=(-12.8\pm 1.1)^{\circ } $.
Therefore the last two fits were performed in mixed propagator approach
(Eq.(\ref{mth:mixing})). Mixing self-energy $ \Pi _{\rho \omega } $
was taken as a free parameter for the model~3 and calculated from
the world average $ B(\omega \rightarrow 2\pi ) $ for the model~4.
The model~4 was used to estimate the contribution from higher vector
resonances $ \rho ' $, $ \omega ' $, which was introduced as
a pure real parameter $ a_{\pi ^{0}\gamma } $. The fitted value
of $ a_{\pi ^{0}\gamma } $ is compatible with zero. All the models~$ 2-4 $
describe the experimental data equally well.

\subsection{Systematic errors.}

Systematic error contributions for obtained cross section parameters
are summarized in the Table \ref{tab:syserror}.

Systematic error of luminosity determination originates mostly from
inaccuracy in cross section calculation ($ 1\% $) and uncertainty
of detection efficiency of the luminosity process, which was estimated
using different angle and acollinearity selection cuts. The total
error of the integrated luminosity determination was $ \sim 2-3\% $.

Primary selection efficiency depends on simple kinematic cuts and
is independent of c.m.s. energy. Thus, its systematic error emerging
from simulation inaccuracy was studied by comparison of simulated
and experimental event distributions at $ \omega  $-resonance peak,
where backgrounds are negligible. Systematic error of primary selection
efficiency does not exceed $ 1.5\% $. 
\begin{table}[h!]

\caption{\label{tab:syserror}Contributions to the systematic errors of the
cross section parameters }

\centering 

\begin{tabular}{|l|c|c|c|}
\hline 
Source&
 $ \sigma _{\omega \pi ^{0}\gamma } $&
 $ \sigma _{\rho ^{0}\pi ^{0}\gamma } $&
 $ \varphi  $\\
\hline
Integrated luminosity &
 $ 2.0\% $&
 $ 3.1\% $&
 $ 8\% $\\
\hline
Three photons selection efficiency &
 $ 1.5\% $&
 $ 1.5\% $&
 $ 4\% $\\
\hline
Final selection efficiency &
 $ 1.6\% $&
 $ 5.6\% $&
 $ 22\% $\\
\hline
Additional clusters &
 $ 0.3\% $&
 $ 1\% $&
$ 2\% $\\
\hline
PDG table errors &
 $ 0.1\% $&
 $ 4\% $&
 $ 4\% $\\
\hline
Total (no model error)&
 $ 3.0\% $&
 $ 9.0\% $&
 $ 24\% $ \\
\hline
\end{tabular}
\end{table}

The machine background changed with c.m.s. energy. To study its influence
on detection efficiency we merged recorded experimental background
events with the simulated ones. Comparison of detection efficiencies
obtained by simulation with and without merging machine background
gives an estimate of the detection efficiency error from this source,
not exceeding $ 0.5\% $.

The final class \textbf{\textsl{A}} selection conditions contain cuts
in invariant masses and complex kinematic parameters $ \chi ^{2}_{3\gamma } $,
$ \chi ^{2}_{2\gamma } $. Dependences of their efficiencies on
c.m.s. energy and ISR photon energy in experiment and simulation may
differ. In order to evaluate systematic error coming from this source,
approximations were done with different cuts in these parameters.

Substantial systematic error contributions to the $ \sigma _{\rho ^{0}\pi ^{0}\gamma } $
and $ \varphi  $ come from inaccuracy of PDG data, mostly from
the $ \Gamma _{\omega } $ uncertainty. 
\begin{table}[h!]

\caption{\label{tab:points}\protect$ e^{+}e^{-}\rightarrow \pi ^{0}\gamma \protect $
cross section. \protect$ \delta E\protect $ is a c.m.s. energy
spread, \protect$ IL\protect $ is an integrated luminosity, \protect$ N\protect $
is a number of events, \protect$ N_{bg}\protect $ is an estimated
number of background events, \protect$ \varepsilon _{\pi ^{0}\gamma }\protect $
is detection efficiency (Eq.(\ref{mth:viseff})) of the process \protect$ e^{+}e^{-}\rightarrow \pi ^{0}\gamma \protect $,
\protect$ \beta _{\pi ^{0}\gamma }\protect $ is a factor taking
into account radiative correction (Eq.(\ref{mth:visbeta})) and energy
spread. Energy (\protect$ \sqrt{s}\protect $) is corrected according
to fitted \protect$ \Delta E\protect $, its error is \protect$ 0.08\protect $~MeV.
The first error of the cross section \protect$ \sigma _{\pi ^{0}\gamma }\protect $
is statistical, the second one is systematic.}

{\footnotesize
\begin{tabular}{|c|c|l|c|c|c|c|l|}
\hline
$ \scriptstyle \sqrt{s} ${\tiny , MeV}&
$ \scriptstyle \delta E ${\tiny , MeV}&
$ \scriptstyle L, nb^{-1} $&
$ \scriptstyle N $&
$ \scriptstyle N_{bg} $&
$ \scriptstyle \varepsilon _{\pi ^{0}\gamma } $&
$ \scriptstyle \beta _{\pi ^{0}\gamma } $&
$ \scriptstyle \sigma _{\pi ^{0}\gamma}, nb $\\
\hline
$  599.52 $&
$  0.14 $&
$  39.90\pm 0.30 $&
$  60 $&
$  46.0 $&
$  0.315 $&
$  0.908 $&
$  1.23\pm 0.77\pm 0.19 $\\
$  629.51 $&
$  0.15 $&
$  46.09\pm 0.33 $&
$  62 $&
$  44.8 $&
$  0.316 $&
$  0.904 $&
$  1.31\pm 0.68\pm 0.43 $\\
$  659.52 $&
$  0.16 $&
$  40.02\pm 0.33 $&
$  37 $&
$  32.0 $&
$  0.313 $&
$  0.899 $&
$  0.45\pm 0.63\pm 0.17 $\\
$  689.56 $&
$  0.19 $&
$  48.31\pm 0.38 $&
$  48 $&
$  33.0 $&
$  0.316 $&
$  0.893 $&
$  1.10\pm 0.59\pm 0.30 $\\
$  719.51 $&
$  0.18 $&
$  58.43\pm 0.43 $&
$  69 $&
$  36.0 $&
$  0.323 $&
$  0.886 $&
$  1.97\pm 0.56\pm 0.18 $\\
$  749.50 $&
$  0.20 $&
$  50.90\pm 0.42 $&
$  84 $&
$  26.8 $&
$  0.317 $&
$  0.866 $&
$  4.09\pm 0.73\pm 0.28 $\\
$  759.50 $&
$  0.20 $&
$  41.88\pm 0.39 $&
$  107 $&
$  20.7 $&
$  0.316 $&
$  0.846 $&
$  7.71\pm 1.02\pm 0.42 $\\
$  763.50 $&
$  0.21 $&
$  38.80\pm 0.38 $&
$  124 $&
$  18.7 $&
$  0.317 $&
$  0.834 $&
$  10.27\pm 1.19\pm 0.47 $\\
$  769.50 $&
$  0.21 $&
$  43.60\pm 0.40 $&
$  234 $&
$  20.3 $&
$  0.319 $&
$  0.812 $&
$  18.95\pm 1.45\pm 1.08 $\\
$  773.50 $&
$  0.21 $&
$  62.77\pm 0.48 $&
$  531 $&
$  28.7 $&
$  0.319 $&
$  0.794 $&
$  31.62\pm 1.51\pm 1.45 $\\
$  777.50 $&
$  0.21 $&
$  76.73\pm 0.53 $&
$  1544 $&
$  35.0 $&
$  0.319 $&
$  0.776 $&
$  79.60\pm 2.13\pm 2.15 $\\
$  778.71 $&
$  0.22 $&
$  6.88\pm 0.16 $&
$  162 $&
$  3.2 $&
$  0.319 $&
$  0.772 $&
$  93.89\pm 8.13\pm 3.82 $\\
$  779.48 $&
$  0.24 $&
$  43.39\pm 0.39 $&
$  1282 $&
$  20.0 $&
$  0.319 $&
$  0.770 $&
$  118.49\pm 3.46\pm 2.94 $\\
$  780.59 $&
$  0.23 $&
$  132.36\pm 0.68 $&
$  4989 $&
$  61.4 $&
$  0.319 $&
$  0.772 $&
$  151.36\pm 2.20\pm 4.40 $\\
$  781.63 $&
$  0.24 $&
$  351.62\pm 1.10 $&
$  15259 $&
$  164.1 $&
$  0.319 $&
$  0.779 $&
$  172.82\pm 1.43\pm 5.15 $\\
$  782.52 $&
$  0.21 $&
$  81.11\pm 0.53 $&
$  3523 $&
$  37.8 $&
$  0.319 $&
$  0.793 $&
$  169.83\pm 2.94\pm 4.76 $\\
$  783.51 $&
$  0.21 $&
$  74.90\pm 0.51 $&
$  3150 $&
$  34.7 $&
$  0.319 $&
$  0.816 $&
$  159.53\pm 2.93\pm 5.46 $\\
$  785.51 $&
$  0.22 $&
$  73.73\pm 0.51 $&
$  2391 $&
$  33.3 $&
$  0.320 $&
$  0.883 $&
$  113.15\pm 2.39\pm 3.18 $\\
$  789.50 $&
$  0.22 $&
$  56.91\pm 0.46 $&
$  930 $&
$  24.6 $&
$  0.320 $&
$  1.044 $&
$  47.61\pm 1.66\pm 1.43 $\\
$  793.49 $&
$  0.23 $&
$  53.03\pm 0.45 $&
$  456 $&
$  22.4 $&
$  0.320 $&
$  1.201 $&
$  21.26\pm 1.10\pm 0.80 $\\
$  799.49 $&
$  0.23 $&
$  51.86\pm 0.45 $&
$  285 $&
$  21.3 $&
$  0.320 $&
$  1.411 $&
$  11.27\pm 0.77\pm 0.31 $\\
$  809.49 $&
$  0.25 $&
$  65.73\pm 0.52 $&
$  189 $&
$  25.8 $&
$  0.318 $&
$  1.660 $&
$  4.70\pm 0.43\pm 0.13 $\\
$  819.49 $&
$  0.24 $&
$  115.74\pm 0.70 $&
$  233 $&
$  43.7 $&
$  0.318 $&
$  1.775 $&
$  2.90\pm 0.25\pm 0.19 $\\
$  839.47 $&
$  0.25 $&
$  144.83\pm 0.80 $&
$  179 $&
$  52.6 $&
$  0.320 $&
$  1.711 $&
$  1.59\pm 0.18\pm 0.07 $\\
$  879.45 $&
$  0.27 $&
$  170.26\pm 0.91 $&
$  100 $&
$  50.4 $&
$  0.318 $&
$  1.269 $&
$  0.72\pm 0.16\pm 0.17 $\\
$  919.43 $&
$  0.32 $&
$  327.70\pm 1.32 $&
$  137 $&
$  77.7 $&
$  0.319 $&
$  1.048 $&
$  0.54\pm 0.12\pm 0.04 $\\
$  939.45 $&
$  0.30 $&
$  291.22\pm 1.28 $&
$  99 $&
$  63.0 $&
$  0.318 $&
$  1.007 $&
$  0.39\pm 0.12\pm 0.09 $\\
$  949.45 $&
$  0.29 $&
$  259.10\pm 1.22 $&
$  86 $&
$  53.9 $&
$  0.317 $&
$  0.993 $&
$  0.39\pm 0.13\pm 0.06 $\\
$  957.45 $&
$  0.29 $&
$  241.63\pm 1.18 $&
$  79 $&
$  48.8 $&
$  0.317 $&
$  0.984 $&
$  0.40\pm 0.13\pm 0.06 $\\
$  969.46 $&
$  0.30 $&
$  245.65\pm 1.21 $&
$  84 $&
$  47.6 $&
$  0.319 $&
$  0.969 $&
$  0.48\pm 0.13\pm 0.10 $\\
\hline
\end{tabular}}

\end{table}

\section{Results}

\begin{figure}[h!]
{\centering \resizebox*{0.9\textwidth}{!}{\includegraphics{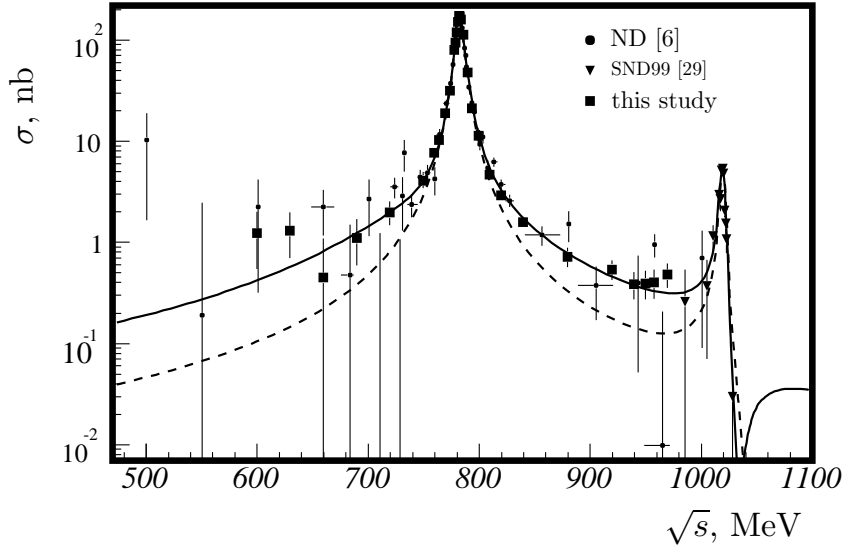}} \par}

\caption{\label{fig:born}\protect$ e^{+}e^{-}\rightarrow \pi ^{0}\gamma \protect $
cross section. Solid line depicts the cross section in the model~2,
dashed line~--- in the model~1. Data from ND experiment are grouped
by energies and shifted taking into account current world average
value of the \protect$ \omega \protect $-meson mass. }
\end{figure}

Our final results are based on model~2 approximation. Differences
in approximation results for models $ 2 $--$ 4 $ were considered
as model error contributions to total systematic errors. As a result
we present: \begin{eqnarray}
\sigma _{e^{+}e^{-}\rightarrow \omega \rightarrow \pi ^{0}\gamma } & = & (155.8\pm 2.7\pm 4.8)\, \mbox {nb},\label{mth:sigom} \\
\sigma _{e^{+}e^{-}\rightarrow \rho ^{0}\rightarrow \pi ^{0}\gamma } & = & (0.58\pm 0.13\pm 0.08)\, \mbox {nb}\label{mth:sigrho} \\
\phi _{\rho \omega } & = & (-10.2\pm 6.5\pm 2.5)\, \mathrm{degrees}\label{mth:phrho} 
\end{eqnarray}
 Detailed point by point listing of the measured $ e^{+}e^{-}\rightarrow \pi ^{0}\gamma  $
cross section is presented in Table \ref{tab:points}. Systematic
error of the experimental cross section is determined by systematic
errors of integrated luminosity, detection efficiency, radiative correction,
and background subtraction. It is worth mentioning that systematic
errors for different c.m.s. energy points are highly correlated. Measured
cross section and data from \cite{cite:Dolinsky:nd,cite:Achasov:phip0g}
are also plotted in Fig.\ref{fig:born}. 

Decay parameters expressed in terms of probabilities and partial widths
are: \begin{eqnarray}
Br(\omega \rightarrow \pi ^{0}\gamma )\cdot B(\omega \rightarrow e^{+}e^{-}) & = & (6.50\pm 0.11\pm 0.20)\times 10^{-6},\label{mth:bromee} \\
Br(\rho \rightarrow \pi ^{0}\gamma )\cdot B(\rho \rightarrow e^{+}e^{-}) & = & (2.34\pm 0.53\pm 0.33)\times 10^{-8},\label{mth:brrhoee} \\
Br(\omega \rightarrow \pi ^{0}\gamma ) & = & (9.34\pm 0.15\pm 0.31)\, \%,\label{mth:brom} \\
Br(\rho ^{0}\rightarrow \pi ^{0}\gamma ) & = & (5.15\pm 1.16\pm 0.73)\times 10^{-4},\label{mth:brrho} \\
\Gamma _{\omega \rightarrow \pi ^{0}\gamma } & = & 788\pm 12\pm 27\, \mbox {keV},\label{mth:gom} \\
\Gamma _{\rho ^{0}\rightarrow \pi ^{0}\gamma } & = & 77\pm 17\pm 11\, \mbox {keV}\label{mth:grho} 
\end{eqnarray}

Obtained results statistically agree with previous measurements. The
partial width $ \Gamma _{\rho ^{0}\rightarrow \pi ^{0}\gamma } $
is in a good agreement with the world average $ \Gamma _{\rho ^{\pm }\rightarrow \pi ^{\pm }\gamma } $.
Phenomenological estimates using various models \cite{cite:Geffen:thnqm,cite:ODonnell:thvdm,cite:Singer:thcbm,cite:Barik:theff,cite:Zhu:thsum}
do not contradict our result.

The ratio of the partial widths of the $ \omega  $,$ \rho \rightarrow \pi ^{0}\gamma  $
decays required by strict $ SU(3) $ symmetry \cite{cite:ODonnell:thvdm}
is equal to $ 9.47 $, which agrees with our measurement: \begin{eqnarray}
\frac{\Gamma _{\omega \rightarrow \pi ^{0}\gamma }}{\Gamma _{\rho ^{0}\rightarrow \pi ^{0}\gamma }} & = & 10.3\pm 2.5\pm 1.4,\label{mth:rgomrho} 
\end{eqnarray}

\section{Conclusions}

The most accurate measurement of $ e^{+}e^{-}\rightarrow \pi ^{0}\gamma  $
cross section is performed at the c.m.s. energy region of $ 0.60-0.97 $~GeV
at the VEPP-2M collider with the SND detector. At present experimental
accuracy level this cross section is well described by vector meson
dominance model taking into account transitions $ \phi ,\omega ,\rho \rightarrow \pi ^{0}\gamma  $.
In this model the cross sections of the processes $ e^{+}e^{-}\rightarrow \omega \rightarrow \pi ^{0}\gamma  $
and $ e^{+}e^{-}\rightarrow \rho ^{0}\rightarrow \pi ^{0}\gamma  $
at corresponding meson masses are measured. Partial widths, their
rations and decay probabilities of corresponding decays were evaluated.
Results are presented in Eqs.(\ref{mth:sigom})-(\ref{mth:rgomrho})
and in the Table \ref{tab:points}.

Measured values of the $ \omega ,\rho ^{0}\rightarrow \pi ^{0}\gamma  $
decay parameters are consistent with earlier experimental results.
Partial width of the $ \rho ^{0}\rightarrow \pi ^{0}\gamma  $~decay
is in a good agreement with that of $ \rho ^{\pm }\rightarrow \pi ^{\pm }\gamma  $~decays.
These values also do not contradict various phenomenological estimations.
Obtained value of the $ \rho -\omega  $ interference phase could
be well explained by electromagnetic $ \rho -\omega  $ mixing.
Our results have higher accuracy than the world average for $ \rho ^{0}\rightarrow \pi ^{0}\gamma  $
and $ \omega \rightarrow \pi ^{0}\gamma  $.

\section{Acknowledgments}

This work was partially supported with RFBR grants \mbox{00-15-96802},
\mbox{01-02-16934-a}, and grant No.78~1999 of Russian Academy of
Science for young scientists.

\end{document}